\documentclass[aps,twocolumn,groupedaddress,superscriptaddress,floatfix,amsmath,amssymb,prl]{revtex4-2}
\usepackage{amsmath}
\usepackage{amssymb}
\usepackage{graphicx}
\usepackage{textcomp}
\usepackage{bm}

\usepackage{booktabs}
\usepackage{siunitx}

\usepackage[usenames,dvipsnames]{color}	

\usepackage[colorlinks,citecolor=Blue,linkcolor=Red, urlcolor=Blue]{hyperref}

\begin{document}
\title{Quantum hardware simulating \\ four-dimensional inelastic neutron scattering}
    
\author{A. Chiesa\footnote{These authors contributed equally to this work.}}    
\affiliation{Dipartimento di Scienze Matematiche, Fisiche e Informatiche, Universit\`a di Parma, I-43124 Parma, Italy}    
\author{F. Tacchino$^*$}
\affiliation{Dipartimento di Fisica, Universit\`a di Pavia, via Bassi 6, I-27100 Pavia, Italy}
\author{M. Grossi}
\affiliation{Dipartimento di Fisica, Universit\`a di Pavia, via Bassi 6, I-27100 Pavia, Italy}
\affiliation{IBM Italia s.p.a., Circonvallazione Idroscalo, 20090 Segrate (MI), Italy}
\author{P. Santini}    
\affiliation{Dipartimento di Scienze Matematiche, Fisiche e Informatiche, Universit\`a di Parma, I-43124 Parma, Italy}    
\author{I. Tavernelli}
\affiliation{IBM Research, Zurich Research Laboratory, Zurich, Switzerland}
\author{D. Gerace}
\affiliation{Dipartimento di Fisica, Universit\`a di Pavia, via Bassi 6, I-27100 Pavia, Italy}
\author{S. Carretta}
\email{stefano.carretta@unipr.it}    
\affiliation{Dipartimento di Scienze Matematiche, Fisiche e Informatiche, Universit\`a di Parma, I-43124 Parma, Italy}    

\date{\today}

\maketitle

{\bf  Finite-size spin systems are test-beds for quantum phenomena \cite{Gaudenzi} and could constitute key elements in future spintronics devices \cite{WW2012,WW2013,Wegewijs,Cervetti,Loth2015}, long-lasting nano-scale memories \cite{Dycene} or noise-resilient quantum computing platforms \cite{Hill,modules,WW2014,jacsYb}. 
Inelastic Neutron Scattering is the technique of choice to model these systems, enabling an atomic-scale characterization of the molecular eigenstates \cite{NatPhys12,Mn12,NatComm17, Cr7Co}. 
However, the full potential of molecular magnetism is still largely unexploited. Indeed, while large molecules can be controllably synthesized \cite{Ringofrings,Fe30prl,Fe30ext,Powell}, the exponential scaling of the required resources on a classical computer precludes the simulation of their dynamics and the interpretation of spectroscopic measurements.\\ 
Here we show that quantum computers \cite{Martinis2015,Gambetta2015,Gambetta2017, eigensolver} can efficiently solve this issue.
By simulating prototypical spin systems on the IBM quantum hardware \cite{eigensolver}, we extract dynamical correlations and the associated magnetic neutron cross-section.
We identify the main gating errors and show the potential scalability of our approach. 
The synergy between developments in neutron scattering and quantum processors will enable a big step forward in the design of spin clusters for fundamental and technological applications.}\\
In order to understand the spin dynamics from Inelastic Neutron Scattering (INS) experiments, we need to compute the magnetic neutron cross-section ($T=0$) as a function of the transferred energy ($E$) and momentum (${\bf Q}$) \cite{Lovesey}:
\begin{eqnarray} \nonumber
I({\bf Q},E) &\propto& \sum_{i,j} F_i(Q) F_j^*(Q) \sum_{\substack{\alpha,\beta=\\x,y,z}} \sum_p \left( \delta_{\alpha,\beta}-\frac{Q_\alpha Q_\beta}{Q^2}\right) \\ 
  && \langle 0 | s_i^\alpha | p \rangle \langle p |s_j^\beta| 0\rangle e^{-i {\bf Q}\cdot{{\bf R}_{ij}}} \delta(E- E_p).
\label{eq:crossS}
\end{eqnarray}
Here $F_i (Q)$ is the (known) magnetic form factor for ion $i$, $|0\rangle$ and $|p\rangle$ are the ground and excited molecular eigenstates with energies $E_0=0$ and $E_p$ and ${\bf R}_{ij}$ are the relative positions of the magnetic ions with spin components $s_i^\alpha$. 
\begin{figure*}[t]
	\centering
	\includegraphics[width=0.92\textwidth]{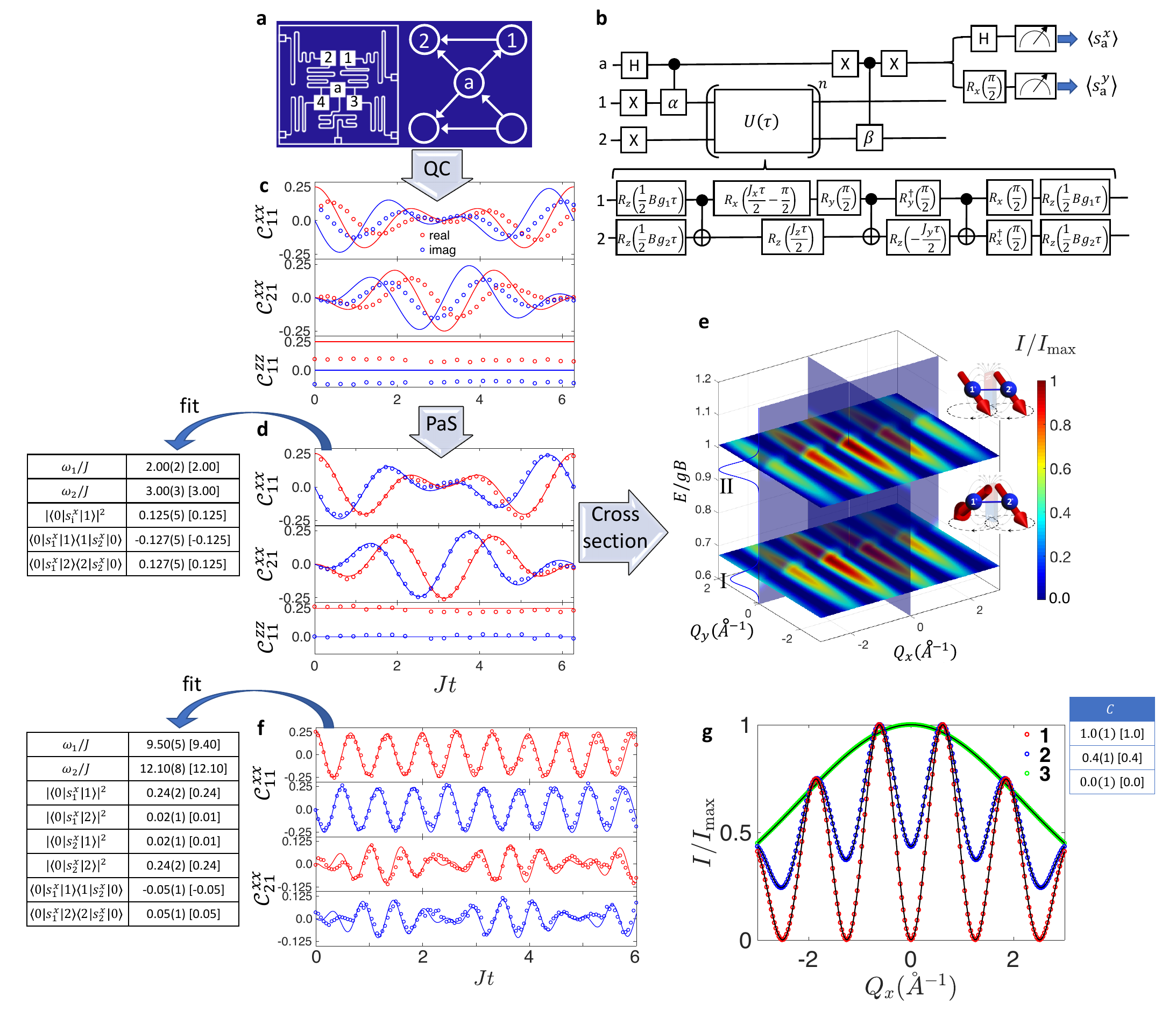} 
	\caption{{\bf Correlation functions and INS cross-section for spin dimers.} {\bf a}, Representation of the ibmqx4 chip and of the connectivity of the five qubits, with arrows pointing towards the target qubit of CNOT gates. {\bf b}, Quantum circuit exploiting the ancilla (a) for computing $\mathcal{C}_{21}^{\beta\alpha}(t) = \left(\langle s^x_{\rm a} \rangle + i \langle s^y_{\rm a} \rangle\right)/2$  ($\alpha,\beta=x,y,z$) \cite{Somma}  (see Methods).
		$U(\tau)$ is the time evolution operator for each elementary time step $\tau=t/n$ of the second-order Suzuki-Trotter decomposition. The two-body evolution is decomposed into a sequence of single-qubit rotations $R_\alpha(\vartheta)={\rm exp}(-i \vartheta s^\alpha)$ and of three CNOTs \cite{PRAdeco}. 
		{\bf c}, Raw dynamical correlations for molecule {\bf 1} calculated on ibmqx4 QC (circles) with $Bg=3J$, compared to the exact dynamics (lines).
		{\bf d}, Phase-and-scale (PaS) corrected dynamical correlations for {\bf 1} and corresponding fit (continuous lines) using Eq. \ref{correl}, with coefficients listed in the table. Exact results are reported in squared brackets, in excellent agreement.
		{\bf e}, Inelastic neutron scattering cross section $I(Q_x,Q_y,Q_z=0,E)$ calculated with the above coefficients, using the form factor and typical inter-atomic distance ($5~\AA$ along $x$ axes) of the dicopper diporphyrines reported in \cite{dicopper}.
		{\bf f}, Pas-corrected Correlation functions for molecule {\bf 2} ($Bg_1=10J$, $Bg_2=12.5J$), calculated on ibmqx4 chip (circles). The coefficients obtained from the fit (lines) are in very good agreement with exact results (see table).
		{\bf g}, $I(Q_x, Q_y=0,Q_z=0, E=E_1)$ cuts, fitted with the analytical expression reported in the Methods (black line) to obtain the concurrence (table).}
	\label{fig1}
\end{figure*}
The excitation energies $E_p$ and the products of spin matrix elements $\langle 0 | s_i^\alpha | p \rangle \langle p |s_j^\beta| 0\rangle$ are the Fourier frequencies and coefficients of dynamical spin-spin correlations functions:
\begin{equation}
\mathcal{C}_{ij}^{\alpha\beta}(t) = \langle s_i^\alpha (t) s_j^\beta \rangle_0 = \sum_p \langle 0 | s_i^\alpha | p \rangle \langle p | s_j^\beta| 0\rangle e^{-i E_p t}.
\label{correl}
\end{equation}
These are the key ingredients for computing $I({\bf Q},E)$ and constitute the hard task for classical computers. Indeed, the calculations of $\mathcal{C}_{ij}^{\alpha\beta}(t)$ for many interesting molecules is presently unfeasible. 
Here we propose the following strategy: (i) use the quantum computer (QC) to simulate $\mathcal{C}_{ij}^{\alpha\beta}(t)$; (ii) extract the excitation energies and products of matrix elements by fitting $\mathcal{C}_{ij}^{\alpha\beta}$ or by performing a classical Fourier transform; (iii) calculate the neutron cross-section on a classical computer by combining the coefficients obtained in (ii) with known quantities such as form factors and position of the ions. 
The simulation of dynamics 
on classical computers is limited to few dozens of spins. 
Conversely, such simulation is exponentially (with the number of spins) more efficient on a QC \cite{NoriRMP}. 
Thanks to the speedup of a quantum processor, the procedure can be repeated many times for different sets of Hamiltonian parameters, thus providing a method to fit INS data of complex spin systems.
\begin{figure*}[t]
	\centering
	\includegraphics[width=0.8\textwidth]{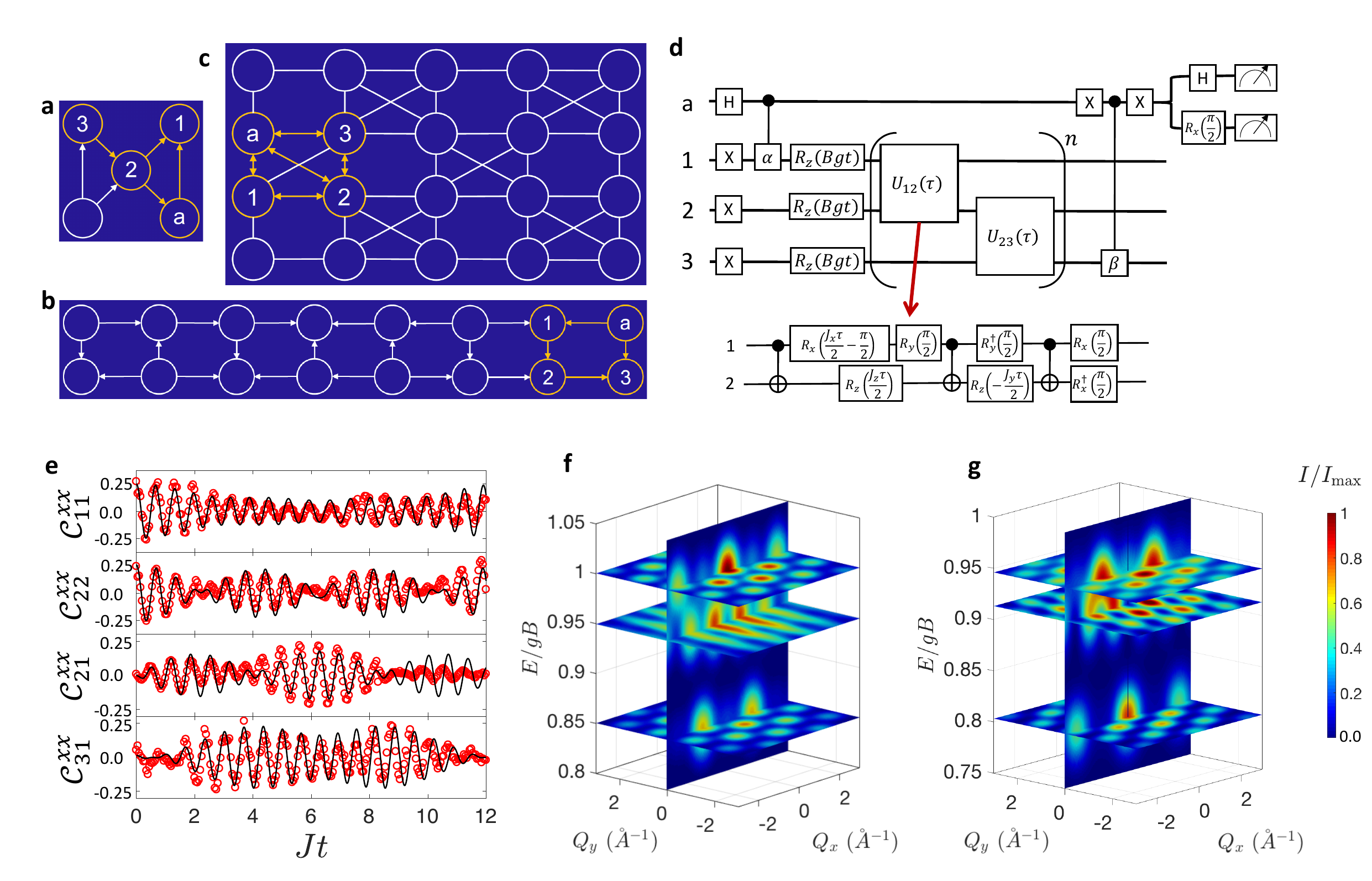} 
	\caption{{\bf Dynamical correlation functions and INS spectrum of spin trimers.} {\bf a} Mapping of the spin trimers on ibmxq4 ({\bf a}), ibmqx5 ({\bf b})  and ibmq20 ({\bf c}) chips, with qubits 1-3 encoding the spin ions and an ancilla (a) for measurements (orange circles). Due to their different topology, ibmqx4 is used for nearest-neighbor correlations $\mathcal{C}_{21}^{\alpha\beta}(t)$, while ibmqx5 for next-to-nearest-neighbor $\mathcal{C}_{31}^{\alpha\beta}(t)$. The improved connectivity and reduced gating errors on the new ibmq20 allows us to implement bi-directional CNOT gates as well as to simultaneously probe nearest-neighbor and next-to-nearest-neighbor correlations on the same device.
		{\bf d}, Quantum circuit to compute $\mathcal{C}_{31}^{\beta\alpha}(t)$, with $Bg=10J$ (molecule {\bf 4}) and $Bg=10J_1^x$ (molecule {\bf 5}). Trotter decomposition is required, due to non-commuting terms $s_1^\alpha s_2^\alpha$ and $s_2^\beta s_3^\beta$, leading to evolution operators $U_{12}(\tau)=e^{-i \tau \sum_{ \alpha}  J_1^\alpha s_1^\alpha s_2^\alpha}$ and $U_{23}(\tau)=e^{-i\tau \sum_{ \alpha}  J_2^\alpha s_2^\alpha s_3^\alpha}$ for each time step $\tau=t/n$. 
		{\bf e}, Dynamical auto-correlations (real part) for molecule {\bf 4} on the two inequivalent ions 1 ($\mathcal{C}_{11}^{xx}$) and 2 ($\mathcal{C}_{22}^{xx}$) and cross-correlations between nearest-neighbor ($\mathcal{C}_{21}^{xx}$) and next-to-nearest-neighbor ($\mathcal{C}_{31}^{xx}$) spins, fitted with a superposition of three frequencies. 
		{\bf f}-{\bf g}, $I(Q_x,Q_y,Q_z=0,E)$ for molecules {\bf 4}-{\bf 5} (equilateral triangles of Cu$^{2+}$ ions with edge $5~\AA$) with slices at the energy of the three peaks.  
		See Supplementary Information for a list of the fitted Fourier frequencies and coefficients, in good agreement with exact results.}
	\label{fig2}
\end{figure*}\\
We experimentally test this method by computing the INS cross-section for prototype spin clusters on IBM chips and address scalability.
Experiments are performed on 5- (ibmqx4), 16- (ibmqx5) and 20-qubits (ibmq20) superconducting processors \cite{PRABlais} composed of fixed-frequency Josephson-junction-based transmon qubits \cite{PRLgates}. Qubit control and readout are achieved using individual superconducting coplanar waveguides (CPW), while another set of CPW resonators (quantum buses), organized as in Fig. \ref{fig1}{\bf a} (for ibmqx4) provide the necessary inter-qubit connectivity. Qubits are cooled to 25 mK in a dilution refrigerator and thus initialized in their ground state (see Methods). All the experiments were run with a large number of measurements (8192) to reduce noise.\\
The benchmark molecules are characterized by the Hamiltonian ($N=2, 3, 4$):
\begin{equation}
\mathcal{H} =  \sum_{\substack{i=1 \\ \alpha=x,y,z}}^{N-1}  J_i^\alpha s_i^\alpha s_{i+1}^\alpha + B \sum_{i=1}^{N}  g_i s_i^z .
\label{Ham}
\end{equation}
We first introduce the simple $N=2$ case to illustrate our approach.
As recently demonstrated \cite{NatComm17}, the four-dimensional (4D)-INS approach allows one to quantify entanglement in effective spin dimers. This is achieved by applying a sizable magnetic field ($B>>J_i^{x,y}$), such that the ground state is factorized and can be exploited as a reference to investigate entanglement in the excited states. Indeed, modulations in the $I({\bf Q})$ of each transition directly reflect the concurrence of the corresponding excited state (see below). Hence, we exploit the ibmqx4 chip to compute the neutron cross-section of spin dimers characterized by different degrees of entanglement, with parameters: $J_i^\alpha=J$, $g_i=g$ (molecule {\bf 1}), $J_i^\alpha=J$, $g_1\neq g_2$ (molecule {\bf 2}) and $J_i^{x,y}=0, J_i^z=J, g_1\neq g_2$ (molecule {\bf 3}). \\
The quantum circuit used to compute dynamical correlation functions is reported in Fig. \ref{fig1}{\bf b} and exploits an ancillary qubit ("a") to measure correlations between logical qubits 1 and 2 \cite{Somma} (see Methods). 
Real (red circles) and imaginary (blue) parts of $\mathcal{C}_{ij}^{\alpha\beta}(t)$ evaluated with ibmqx4 chip for molecule {\bf 1} are reported in Fig. \ref{fig1}{\bf c}. 
The comparison between experimental raw data and the exact evolution (lines) shows a clear overall attenuation and a phase error, mixing real and imaginary parts, mainly originating from systematic coherent errors discussed below.  
As shown in panel {\bf d}, simple and general conditions on $\mathcal{C}_{ii}^{\alpha\alpha}(0)$ allow us to systematically fix these discrepancies (Methods). 
The phase of $\mathcal{C}_{ii}^{\alpha\alpha}(0)$ is corrected by imposing it a real positive value 
and then the same phase correction is applied to the whole time domain. 
To restore the correct amplitude, we apply a circuit-depth dependent scaling factor, obtained from the general sum rule $\langle s_i^2 \rangle = s_i (s_i+1) = \sum_{\alpha} \mathcal{C}_{ii}^{\alpha\alpha}(0)$. 
After application of this phase-and-scale (PaS) procedure, the correct dynamics is perfectly recovered (Fig. \ref{fig1}{\bf d}). Energies $E_p$ and coefficients $\langle 0 | s_i^\alpha | p \rangle \langle p |s_j^\beta| 0\rangle$ entering the cross-section are then extracted by fitting the time-dependence with a combination of $e^{-i E_p t}$ terms (Eq. \ref{correl}) (see Methods). While $z-z$ correlations are constant, we find that $x-x$ ones have two Fourier components. The extracted coefficients are reported in the table, in excellent agreement with exact values (squared brackets). 
The resulting neutron cross-section $I(Q_x,Q_y,Q_z=0,E)$ is shown in Fig. \ref{fig1}{\bf e} (see Methods). 
Correlations evaluated with the IBM chip for molecule {\bf 2} are shown in Fig. \ref{fig1}{\bf f}, after application of our PaS correction. Non-commuting terms of the target Hamiltonian (due to $g_1\neq g_2$) imply the use of a Trotter decomposition, with $n=2$ up to $Jt=2.0$ and $n=4$ for $2.0<Jt\le6.0$ (see Supplementary Information). Although this leads to a much longer circuit (64 gates in total), the agreement between measured and calculated Fourier coefficients (see table) is still very good. Experiments performed with the parameters of cluster {\bf 3} lead to completely mono-chromatic oscillations of the auto-correlation functions and negligible cross-correlations (see Supplementary Information).
The ${\bf Q}$-dependence of the neutron spectra obtained by the present experiments allows us to extract the concurrence $C$, a measure of entanglement \cite{NatComm17} (see Methods). A clear decrease in the amplitude of the $I(Q_x)$ modulations (Fig. \ref{fig1}{\bf g}) fingerprints a decrease of entanglement. Indeed, we obtain $C=1.0(1),0.4(1),0.0(1)$ for molecule {\bf 1}, {\bf 2}, {\bf 3}, in agreement with exact calculations. 
\begin{figure}[b]
	\centering
	\includegraphics[width=0.5\textwidth]{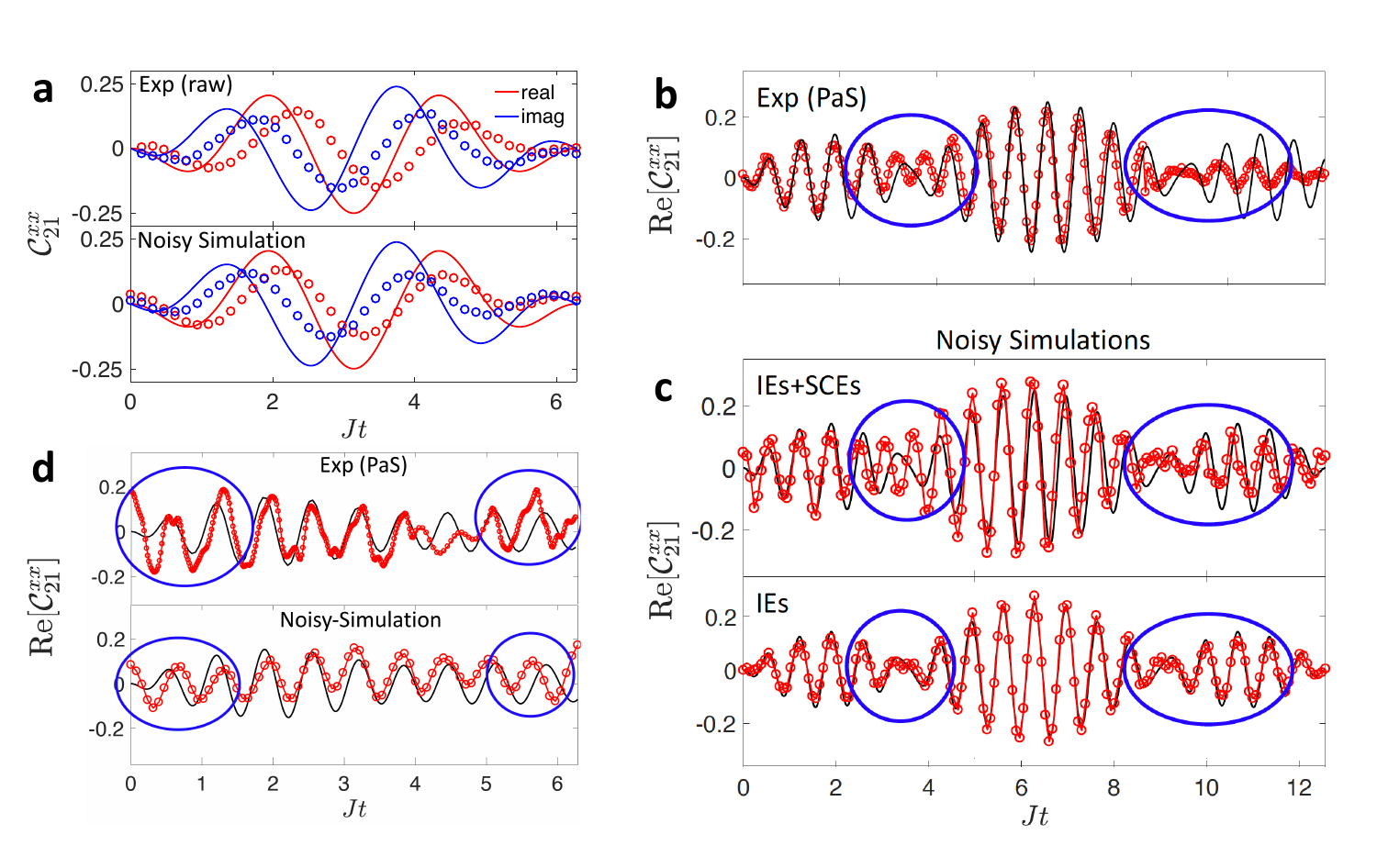} 
	\caption{{\bf Error analysis.} Comparison between simulated (including noise) and experimental (on ibmqx4 chip) dynamical correlation functions for a spin chain of variable length $N$, characterized by the target spin Hamiltonian $\mathcal{H} =  J \sum_{i=1}^{N-1}  {\textbf s}_i \cdot {\textbf s}_{i+1} + g B \sum_{i=1}^{N}  s_i^z$. {\bf a}, raw data for molecule {\bf 1} ($N=2$), with attenuation and phase error well reproduced by our errors-model. {\bf b}, PaS-corrected data for molecule {\bf 4} ($N=3$), compared to noisy simulations ({\bf c}) with both SCEs and IEs or only IEs. Blue circles highlight residual discrepancies (not recovered by PaS) from the exact evolution. Clearly, residual errors are due to SCEs. {\bf d} Extension to $N=4$.}
	\label{fig3}
\end{figure}\\
In order to show the effectiveness of our scheme on more complex Hamiltonians, we apply it to isotropic (molecule {\bf 4}, $J_i^x=J_i^y=J_i^z=J$, $g_i=g$) and anisotropic (molecule {\bf 5}, $J_i^x=J_i^y = -0.5J_i^z$) spin trimers. To further lower the symmetry, in the latter we assume two different exchange bonds ($J_2^\alpha = 0.7 J_1^\alpha$). 
For the calculations we exploit two chips which, thanks to the different coupling topology, allow us to probe both nearest-neighbors (on ibmqx4, Fig. \ref{fig2}{\bf a}) and next-to-nearest-neighbors correlation functions (on ibmqx5, Fig. \ref{fig2}{\bf b}). 
We also employed the new and optimized ibmq20 chip (Fig. \ref{fig2}{\bf c}) to compute correlations for molecule {\bf 5}. Thanks to its improved connectivity, it allows us to probe both nearest-neighbors and next-to-nearest-neighbors correlation functions on the same device.
The quantum circuit is reported in Fig. \ref{fig2}{\bf d}, while
examples of the resulting $\mathcal{C}_{ij}^{xx}$ for molecule {\bf 4} are displayed in panel {\bf e} and in the Supplementary Information.  
We have hereby used a Trotter decomposition with $n=2$, which allows us to extract the correct Fourier coefficients (Tables S2-S3) and to accurately compute the 4D-INS spectrum. The latter is shown in panels {\bf f}-{\bf g} for molecules {\bf 4}-{\bf 5} and is very close to that obtained by diagonalizing $\mathcal{H}$ (Figs. S21-S22). These results are remarkable and show that the propagation of gating errors is well recovered by PaS. \\
To assess the scalability of the scheme, 
we address the various errors and their propagation by comparing targeted experiments on the real hardware to numerical {\it noisy simulations} including all main errors. In particular, we consider systematic-coherent (SCEs), measurement (MEs) and incoherent (IEs) errors including relaxation and dephasing.
By focusing on elementary gates, we quantify all these errors and identify SCEs as the leading ones (see Methods).
Then, to test our error model (Fig. \ref{fig3}), we compare the numerical simulation of dynamical correlation functions (involving a sequence of many noisy gates) to experimental results on 3, 4 and 5 qubits (including the ancilla).  
Different classes of errors can be easily distinguished, since they produce different features on the computed correlations: while IEs essentially yield an overall attenuation in the oscillations of dynamical correlations, the concatenation of SCEs can significantly alter the dynamics. The comparison between our noisy simulation and the raw data for molecule {\bf 1} is shown in Fig. \ref{fig3}{\bf a}. 
Both attenuation and phase errors are well reproduced by our model and mostly fixed by the PaS correction (see Methods). However, residual off-resonant-rotation (ORR) errors \cite{PRAZgates} lead to significant effects with increasing $N$. Some of these are highlighted by blue circles in the experimental data shown in panels {\bf b} ($N=3$) and {\bf d} ($N=4$). Remarkably, these features are well captured by our errors-model. It is also worth noting that such discrepancies originate mainly from SCEs, as clearly demonstrated in Fig. \ref{fig3}{\bf c}: IEs alone yield an overall attenuation. \\
We now quantitatively investigate the scalability of the method for a spin chain, by including in our noisy simulations a realistic propagation of errors. Relaxation and dephasing errors depend on the total gating time. Since the simulation of odd and even bonds can be performed in parallel, our results show that these are not currently hindering the quantum calculation (Methods).
\begin{figure}[t]
	\centering
	\includegraphics[width=0.4\textwidth]{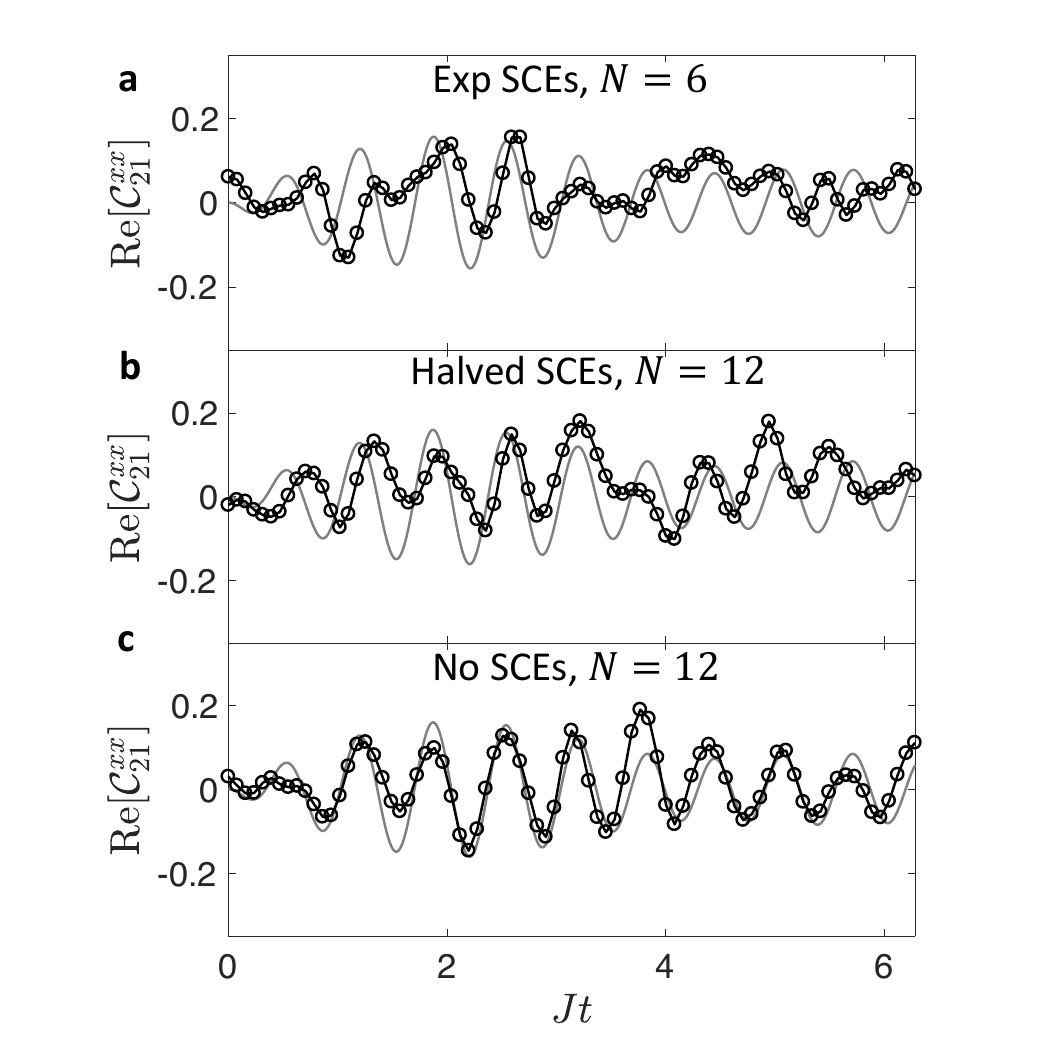} 
	\caption{{\bf Scalability of the approach.} Examples of dynamical correlation functions simulated by assuming $N+1$ qubits (including an ancilla) characterized by the errors determined above ({\bf a}), with halved SCEs ({\bf b}) and with only IEs ({\bf c}). Data were smoothed to reduce statistical noise and treated with our PaS correction. Grey lines are the exact dynamics. The number of Trotter steps in the reported simulations increases as $n=\text{round} [2\sqrt{N-2}]$, in order to keep the second-order digital error fixed at variable $N$ (see Methods). }
	\label{fig4}
\end{figure}
Fig. \ref{fig4} shows some examples of dynamical correlations, numerically simulated by including all errors determined above. 
Using a realistic noise model, 
the maximum $N$ enabling a reliable computation is $N\approx6$ (Fig. \ref{fig4}{\bf a}). A simple halving of SCEs (keeping fixed IEs) should already enable a good simulation with $N=12$ (Fig. \ref{fig4}{\bf b}). 
Fig. \ref{fig4}{\bf c} shows that, by removing the main SCEs while keeping IEs, the simulation for $N=12$ performs very well, making larger calculations possible in the future. 
Hence, SCEs are currently the main limiting factor and their mitigation will enable to scale to an interesting number of qubits, even without improving IEs.
Our approach can be extended with forthcoming technological progresses to a number of spins that 
would make a non error-corrected quantum hardware much more efficient than a classical device for the practical interpretation of many experimental data. 
These results, combined with remarkable developments occurring in neutron scattering facilities, open new avenues in the design and understanding of molecular spin systems.

\vspace{0.5cm}

\noindent
\textbf{Acknowledgements}\\
\begin{small}
	This work was partly funded by the Italian Ministry of Education and Research (MIUR) through PRIN Project 2015 HYFSRT  ``Quantum Coherence in Nanostructures of Molecular Spin Qubits''. 
	Very useful discussions with G. Amoretti and G. Prando are gratefully acknowledged.
\end{small}
\\

\noindent
\textbf{Author contributions}\\
\begin{small}
A.C., F.T. and M.G. used the IBM chips to implement gate sequences. Analysis of the experimental results was made by A.C., F.T., D.G., I.T. and S.C..
Numerical calculations have been performed by A.C. and F.T..
P.S., I.T., D.G. and S.C. conceived the work and discussed the results with other coauthors.
A.C. and S.C. wrote the manuscript with inputs from all coauthors.
\end{small}
\\

\noindent
\textbf{Competing interests}\\
\begin{small}
	The authors declare no competing interests.
\end{small}
\\

\noindent
\textbf{Additional information}\\
\begin{small}
	{\bf Supplementary information} is available for this paper at
\end{small}

\vspace{0.3cm}

\noindent
\textbf{Methods}\\
\begin{small}
	{\bf Experimental implementation of the gates.} 	
	The ibmqx4 5-qubit processor is described schematically in Fig. \ref{fig1}{\bf a}. All 5 transmon qubits are coupled individually with CPW resonators used for both qubit control and readout. Two additional shared CPW resonators are used to couple the central qubit, a, with all other qubits (1 to 4) as well as qubit 2 with 4, and 1 with 3.
	The ibmqx5 quantum processor is composed of 16 transmon qubits organized as depicted in Fig. \ref{fig2}{\bf b}. In this case, 4 qubits (enclosed  by orange circles in Fig \ref{fig2}{\bf b}) are used to compute next-to-nearest neighbors correlations on the trimer, which cannot be directly obtained with the connectivity of the ibmqx4 device of Fig. \ref{fig2}{\bf a}. 
	The new ibmq20\_Tokyo processor consists of 20 transmon qubits organized as schematically shown in Fig. \ref{fig2}{\bf c}. 
	Thanks to its improved connectivity, it allows us to implement bidirectional CNOT gates (double-headed arrows) between several pairs of qubits, and to measure both nearest-neighbors and next-to-nearest-neighbors correlations using the same set of qubits (a, 1, 2, 3 in Fig. \ref{fig2}{\bf c}, enclosed by orange circles). Furthermore, it is optimized to reduce noise in the implementation of the elementary gates (see parameters reported on IBM QX  \href{https://quantumexperience.ng.bluemix.net/qx/devices}{webpage}).\\
	The qubits are controlled by microwave pulses that are sent from the electronics operating at room temperature to the quantum chip through attenuated coaxial lines. 
	Single qubit gates are operated at their specific fixed frequencies 
	as specified on IBM QX  \href{https://quantumexperience.ng.bluemix.net/qx/devices}{webpage}. Two-qubit cross-resonance gates are obtained by driving a selected qubit $Q_c$ (characterized by a fixed frequency $\omega_c$) at the frequency $\omega_t$ of the target qubit. The pairs of control-target qubits are defined by the coupling map in Figs. \ref{fig1}{\bf a} and \ref{fig2}{\bf a}-{\bf c}. 
	Note that in ibmqx4 and ibmqx5 CNOT gates can only be implemented between two connected qubits with a fixed orientation that defines control and target qubits (arrows in Figs. \ref{fig2}{\bf a}-{\bf b}). 
	The state of each qubit is measured at its readout resonator frequency; the reflected readout signals are amplified first by a Josephson parametric converter followed by HEMT amplifiers operating at 4K. \\
	It is important to note that all non-elementary quantum operations need to be decomposed into elementary operations of the fundamental gate set (Bloch sphere rotations and CNOT gates) prior coding them in the IBM processors (see Supplementary Information).\\
	{\bf Quantum circuit for dynamical correlation functions.} 
		The quantum circuit (see Fig. \ref{fig1}{\bf b} for the $N=2$ case) used to measure correlation functions of the form $\langle A^\dagger B\rangle$ (where $A$ and $B$ are unitary operators) exploits an ancillary qubit "a" \cite{Somma}. In this work we compute {\it spin-spin} correlation functions, hence we focus on $A=\sigma^\beta=2s^\beta$  and $B=\sigma^\alpha=2s^\alpha$. Two X gates are used to bring the system, initially cooled with all qubits in 0, to the ground state of the target Hamiltonian $|\downarrow\downarrow\rangle \equiv |11\rangle$. Conversely, an Hadamard (H) gate initializes the ancilla in $\frac{|0\rangle+|1\rangle}{\sqrt{2}}$.
		 Then we make two controlled evolutions: the first applies $B$ to the system if the ancilla is in $\vert 1 \rangle$, the second evolves the system by $A$ if the ancilla is in $\vert 0 \rangle$. Finally, the expectation value of $2 \sigma^+_{\rm a} =  \sigma^x_{\rm a}  + i  \sigma^y_{\rm a}$ gives the result $\langle A^\dagger B\rangle$.
		This is obtained by repeating the computation twice, alternatively rotating the state of the ancilla by $\pi/2$ about $x$ or $y$ axis before measuring in the $z$ basis.
		The computation of {\it dynamical} correlation functions can be done analogously, by including the quantum simulation of the target system time-evolution (represented by the unitary operator $U=e^{-i \mathcal{H} t}$) between the two controlled operations in which the ancilla acts as a control \cite{Somma}. In this way we obtain $\langle 2\sigma^+_{\rm a} \rangle = \langle U^\dagger A U B \rangle \equiv \langle \sigma^\beta(t) \sigma^\alpha \rangle$.\\	
	{\bf PaS correction.} Raw experimental data are treated by means of the PaS correction, which can systematically correct for a phase error in the measured values of $\mathcal{C}_{ij}^{\alpha\beta}(t)$, as well as for an overall attenuation of the experimental oscillations. From Eq. \ref{correl}, we note that $\mathcal{C}_{ii}^{\alpha\alpha}(0)$ corresponds to a sum of squared absolute values. Hence, independently from the target Hamiltonian, we can impose it a real positive value, thus determining the initial phase correction and then extending it to the whole time domain. Conversely, the overall attenuation is fixed by a scale factor increasing with the circuit depth, through the general sum rule $\langle s_i^2 \rangle = s_i (s_i+1) = \sum_{\alpha} \mathcal{C}_{ii}^{\alpha\alpha}(0)$ reported in the text, depending only on the values of the local spin operators. 
	It is finally worth noting that such overall attenuation, as well as a small constant shift in the observed value of $\langle s^z \rangle$ probably due to asymmetric measurement errors, is not relevant for calculating the inelastic cross-section (usually measured in arbitrary units) and that the oscillation frequencies are very well reproduced. \\
	{\bf Fitting dynamical correlation functions.} 
	$\mathcal{C}_{ij}^{\alpha\beta}(t)$ are fitted with a combination of oscillating functions:
	\begin{eqnarray}\nonumber
	\mathcal{C}_{ij}^{\alpha\beta}(t) = \sum_p [A_{ij}^{\alpha\beta}(\omega_p) + i B_{ij}^{\alpha\beta}(\omega_p)] e^{-i\omega_p t}  \\ \nonumber
	= \sum_p \left[ A_{ij}^{\alpha\beta}(\omega_p) \text{cos}~ \omega_p t + B_{ij}^{\alpha\beta}(\omega_p) \text{sin}~ \omega_p t\right] \\ 
	+ i \sum_p \left[ B_{ij}^{\alpha\beta}(\omega_p) \text{cos}~ \omega_p t - A_{ij}^{\alpha\beta}(\omega_p) \text{sin}~ \omega_p t\right].
	\label{eq:methods}
	\end{eqnarray}
	Here we have recast the Fourier coefficients  $\langle 0 | s_i^\alpha | p \rangle \langle p |s_j^\beta| 0\rangle = A_{ij}^{\alpha\beta}(\omega_p) + i B_{ij}^{\alpha\beta}(\omega_p)$ in order to separate the real and imaginary parts of $\mathcal{C}_{ij}^{\alpha\beta}(t)$, corresponding to the second and third line of Eq. (\ref{eq:methods}).
	Some general conditions impose constraints on the parameters of the fit. 
	First, we note that the eigenstates of the Heisenberg and Ising models are known to be real, thus leading to real matrix elements for $s^x$ and $s^z$ and imaginary ones for $s^y$. This yields $B_{ij}^{\alpha\alpha}(\omega_p)=0$, thus reducing the number of parameters. In addition, we have checked by simulating the related time evolution on the chip (see Supplementary Information) that only $\mathcal{C}_{ij}^{\alpha\alpha}(t)$ dynamical correlations contribute to the cross-section ($\mathcal{C}_{ij}^{xz}(t)$ and $\mathcal{C}_{ij}^{yz}(t)$ are identically zero, while $\mathcal{C}_{ij}^{xy}(t)$ and $\mathcal{C}_{ij}^{yx}(t)$ cancel out). Furthermore, we found that $\langle s^z_i(t) s^z_j \rangle$ are independent of time (thus not contributing to the {\it inelastic} cross-section). At last, 
	the results fulfill axial and permutational symmetries of the target Hamiltonian, when present. The former leads to equivalence between $\mathcal{C}_{ij}^{xx}$ and $\mathcal{C}_{ij}^{yy}$ contributions, the latter between $\mathcal{C}_{ij}^{\alpha\alpha}$ and $\mathcal{C}_{ji}^{\alpha\alpha}$. Since $\mathcal{C}_{ij}^{yy}$ involves a larger number of gates and is thus more error prone than $\mathcal{C}_{ij}^{xx}$, leading to more noisy correlation functions, we used only $\mathcal{C}_{ij}^{xx}$ for calculating the final cross-section (see comparison in the Supplementary Information).\\
	Frequencies $\omega_p$ and Fourier coefficients have been extracted from correlations by combining a Fourier analysis with the FMINUIT package.
	In the examined spin dimers, we found that only two frequencies have non-negligible weight, while in the trimer three frequencies are needed to reproduce the measured oscillations of the correlation functions. \\
		{\bf Suzuki-Trotter decomposition of the spin dynamics.} 
	To reduce the total number of gates while keeping the digital error as small as possible, we employ the decomposition reported in Ref. \cite{PRAdeco} for a general two-spin interaction (details are given in the Supplementary Information) and for molecule {\bf 2} a second-order Suzuki-Trotter expansion:
	\begin{equation}\nonumber
	U(\tau)  = e^{-i \mathcal{H}_1 \tau/2}  e^{-i \mathcal{H}_2 \tau} e^{-i \mathcal{H}_1 \tau/2} = e^{-i (\mathcal{H}_1+\mathcal{H}_2) \tau}+ O(\tau^3) ,
	\end{equation}
	with $\mathcal{H}_1 = B \left( g_1 s_1^z +g_2 s_2^z\right)$ and $\mathcal{H}_2= J \mathbf{s}_1 \cdot \mathbf{s}_2$ indicating one- and two-body terms of the target spin Hamiltonian. \\
	{\bf Concurrence from INS spectrum of spin dimers.}
	The two-qubit entanglement can be quantified by means of the {\it concurrence} ($C$) \cite{concurrence}. For pure two-qubit states $|p\rangle = a |00\rangle+b|01\rangle+c|10\rangle+d|11\rangle$, $C= 2|ad-bc|$.  
	In the present cases, we have found $\mathcal{C}_{ij}^{zz}(t)$ independent of time and hence $d=0$ in the excited states corresponding to the two calculated INS peaks.
	By inserting the expression of $|p\rangle$ in Eq. (\ref{eq:crossS}), we get
	\begin{eqnarray}
	I(Q_x) &\propto& F_1(Q_x)F_2^*(Q_x) \left(  \vert b\vert^2+  \vert c\vert^2+  bc^*e^{iQ_xR}+ \text{h.c.}\right) \\ \nonumber
	&=&  F_1(Q_x)F_2^*(Q_x) \\ \nonumber &\times& \left[  \vert b\vert^2+  \vert c\vert^2+2{\rm Re}(bc^*) {\rm cos} ~Q_xR -2{\rm Im}(bc^*) {\rm sin} ~Q_xR \right]
	\label{eq:ent}
	\end{eqnarray}
	where ${\bf R}_{12}=R {\bf x}$. Hence, the parameters $b$ and $c$ (and, consequently $C=2|bc|$) can be obtained from a fit of $I(Q_x)$. We note that the presence of only two peaks in the spectrum of Fig. \ref{fig1}{\bf e} implies $a=0$.\\
{\bf Characterization of errors on elementary gates.}
We start our investigation from single-qubit rotations. In particular, in the IBM hardware $R_z$ rotations are implemented by simply adding the proper phase to subsequent operations, via classical control hardware and software \cite{PRAZgates}. Therefore, such a gate is essentially perfect and has no duration. Hence, we focus on arbitrary $R_{x,y}$ rotations and perform full tomography of the qubit state after the implementation of the (noisy) gate on the real hardware. In order to reduce the number of operations we initialize the qubit in its ground state and implement a generic $R_{x,y} (\vartheta)$ gate. Measurements in the $z$-basis yield the diagonal elements of the final density matrix, while coherences are obtained by performing measurements in the $x$ and $y$ basis (in these latter cases error propagation was also accounted for in our analysis). 
Simulations were performed by ibmq-qasm-simulator \cite{qasmsim}, properly adapted in order to account not only for incoherent noisy channels, but also for an imperfect coherent dynamics in the implementation of the gates (SCEs). 
Experimental data for an $R_x(\vartheta)$ gate on qubit 0 of ibmqx4 chip, with 8192 counts, are reported in the left panels of Fig. S25.
Similar results are obtained also for other qubits on the same chip. As shown in the Supplementary Information, our simulations (right panels of Fig. S25) are able to catch the main discrepancies between the actually implemented gates and the ideal ones by including a systematic tilt of the rotation axis (ORR error). Such error was already reported for similar devices (see, e.g. \cite{PRAZgates}).
It is important to note that none of the incoherent noisy Pauli channels can produce the same effect on the final density matrix. 
Asymmetric measurement errors could lead to a similar behavior in the diagonal elements of the single-qubit density matrix, but would not yield the error propagation observed in our experiments when several rotations are concatenated (see main text). 
Hence, we assumed for simplicity a symmetric measurement error of about 4-5 \% (consistent with data reported on IBM QX  \href{https://quantumexperience.ng.bluemix.net/qx/devices}{webpage}). This choice results in a pessimistic estimate of the scalability of our approach, because part of the error ascribed to ORRs could be due to a single asymmetric measurement error (at the end of the calculation and on a single qubit).
Next, we investigate errors in the implementation of a CNOT gate. In order to fully characterize them (both in the amplitude and phase of the output), we initialize a pair of qubits on the ibmqx4 chip in one of the four factorized states $\frac{\vert00\rangle \pm \vert10\rangle}{\sqrt{2}}$, $\frac{\vert01\rangle \pm \vert11\rangle}{\sqrt{2}}$, which in an ideal implementation would lead to the entangled Bell basis states $\frac{\vert00\rangle \pm \vert11\rangle}{\sqrt{2}}$, $\frac{\vert01\rangle \pm \vert10\rangle}{\sqrt{2}}$. 
It should be noted that state preparation and final tomography require several rotations, each one affected by the SCEs discussed above. 
We find that the inclusion of such errors is already sufficient to satisfactorily reproduce the experimental results for the noisy CNOT. Conversely, typical coherent errors for the cross-resonance gate implemented in this device (ZZ, ZI, IX, reported e.g. in \cite{PRAcross}) do not have an important effect on the final density matrix and are thus excluded from our simulations. Finally, incoherent errors slightly improve the agreement (see examples of comparison between measured and simulated final density matrices in Fig. S26). 
Hence, we have included them with typical values. We list below the error parameters used in our numerical simulations: (i) systematic tilt of the $x$ axis for $R_x(\vartheta)$ gates of $\pi/16$ towards $z$ and $\pi/8$ towards $y$; systematic tilt of the $y$ axis for $R_y(\vartheta)$ gates of $\pi/16$ towards $z$ and $-\pi/12$ towards $x$. (ii) Symmetric measurement errors $p_m=0.05$ (in agreement with data reported online for this chip). We note that readout errors are minimized in our circuit since only a single ancillary qubit must be measured. (iii) Relaxation rate with typical $T_1 = 30 ~\mu s$. (iv) Incoherent Pauli error channels with $p_u=0.002$ and $p_{CNOT}=0.05$ for single- and two-qubit gates, respectively, in line with gate errors reported for this device. These values were also tuned to reproduce the observed attenuation of correlation functions. Since from analysis of the dynamical correlations it was not possible to discriminate the effect of different Pauli channels, we assumed equally distributed probabilities for each of them (depolarizing channel), thus reducing the number of parameters. A Pauli-Z error channel is also used to model pure dephasing with typical $T_2 = 30 ~\mu s$. (v) Thermal population of the excited qubit state $p_1=0.005$, as reported in Ref. \cite{PRAZgates} (practically negligible).\\	
{\bf Scalability.}  To investigate the scalability of the proposed approach, we have considered as a target system an open spin chain with nearest-neighbors Heisenberg exchange interactions and examine dynamical correlations on one edge of the chain as a function of the number of sites, $N$. This choice ensures that the increase of $N$ does not alter too much the observed quantities, and hence allows us to focus only on the propagation of errors. \\
As stated in the main text, relaxation (and dephasing) errors depend on the total gating time $T$, which does not increase for $N>3$. 
Nevertheless, the error on the final state could scale with $N$ in the worst case. Indeed, each qubit in the excited state (whose thermal population is negligible) has a probability 
$e^{-T/T_1}$ to be still excited after a time $T$. The worst-case fidelity (see Ref. \cite{SciRepSimul}) would be obtained by starting from a state with all the qubits in the excited level, leading to a total error probability $(1-e^{-NT/T_1})$. Analogous considerations also hold for pure dephasing, leading to a decay of off-diagonal elements of the single-qubit density-matrix and a consequent linear scaling of the dephasing error in the worst case (see \cite{modules}). 
However, here we focus on two-spin dynamical correlation functions on a factorized ground state, which provides a reference to study entanglement in the excited states \cite{NatComm17}. Our numerical simulations show that in this case the overall attenuation factor does not strongly increase with $N$. 
This can be seen in Fig. S27, where we report the overall attenuation factor induced by $T_1$ and $T_2$ on the oscillations of $\mathcal{C}_{ij}^{xx}$. For simplicity, errors affecting the ancilla in the idle phase were not included in the numerical simulations. \\
This analysis holds if the number $n$ of Trotter steps (and hence the number of gates per site) is fixed. However, in order to mimic the dynamics of the real system, we need to digitalize the time evolution in such a way to preserve the same digital error at variable $N$. By considering (for simplicity) a first-order Suzuki-Trotter decomposition, the digital error is given by  $\epsilon_N = \frac{\tau^2}{2} \left[\mathcal{H}_{odd},\mathcal{H}_{even}\right]$, where $\tau = \frac{t}{n}$ and 
$\mathcal{H}_{odd} = J \sum_i {\textbf s}_{2i-1} \cdot {\textbf s}_{2i}$ and $\mathcal{H}_{even} = J \sum_i {\textbf s}_{2i} \cdot {\textbf s}_{2i+1}$ are the two non-commuting terms of the target Hamiltonian in our one-dimensional chain. Hence 
\begin{equation}\nonumber
\epsilon_N \sim \frac{\tau^2 J^2}{2} (N-2) \left[ {\textbf s}_{1} \cdot {\textbf s}_{2}, {\textbf s}_{2} \cdot {\textbf s}_{3} \right] \propto \frac{t^2}{n^2} (N-2).
\end{equation}
In particular, for a given simulation time $t$ and by fixing this error to that of the trimer ($N=3, n=2$, which ensures a very good determination of the INS spectrum, see SI), we obtain that the number of needed steps slowly increases with the $N$, i.e. $n=\text{round} [2\sqrt{N-2}]$. This condition is applied to all the simulations at increasing $N$. \\
We finally examine the scaling of the total number of gates required to compute the whole neutron cross-section. Indeed, in general we need to calculate correlations between any spin-pair in the target system. Accordingly, the total number of dynamical correlations to be computed scales (polynomially) as $N^2$. 
This, however, does not require to increase the number of ancillae. Indeed, we only need to repeat the calculation on the quantum hardware a polynomial number of times, with slightly different circuits for each spin pair. \\
	\vspace{0.35cm}\\
	{\bf Data availability.} \\
	The data that support the plots and other findings of this study are available from the corresponding author upon reasonable request.\\
	{\bf Code availability.}\\
	The custom Python scripts for the quantum hardware and original codes are available from the corresponding author upon reasonable request.

	
	

\end{small}

\end{document}